\documentclass[preprint,showpacs,aps]{revtex4}

\usepackage{graphicx}
\usepackage{dcolumn}
\usepackage{bm}

\begin{document}

\begin{titlepage}

\title{Huge enhancement of electromechanical responses in
compositionally modulated Pb(Zr$_{1-x}$Ti$_{x}$)O$_{3}$ }

\author{Ningdong Huang, Zhirong Liu, Zhongqing Wu, Jian Wu,
Wenhui Duan\footnote{Author to whom any correspondence should be addressed.}}
\address{Department of Physics, Tsinghua University, Beijing
100084, People's Republic of China}
\author{Bing-Lin Gu}
\address{ Department of Physics, and Center for Advanced Study,
Tsinghua University, Beijing 100084, People's Republic of China}
\author{Xiao-Wen Zhang}
\address{State Key Laboratory of New Ceramics and Fine Processing,
and Department of Materials Science and Engineering, Tsinghua
University, Beijing 100084, People's Republic of China}
\date{\today}

\begin{abstract}
Monte Carlo simulations based on a first-principles-derived
Hamiltonian are conducted to study the properties of
Pb(Zr$_{1-x}$Ti$_{x}$)O$_{3}$ (PZT) alloys compositionally
modulated along the [100] pseudocubic direction near the
morphotropic phase boundary (MPB). It is shown that compositional
modulation causes the polarization to continuously rotate away
from the modulation direction, resulting in the unexpected triclinic
and C-type monoclinic ground states and huge enhancement of
electromechanical responses (the peak of piezoelectric coefficient
is as high as 30000 pC/N). The orientation dependence of
dipole-dipole interaction in modulated structure is revealed as
the microscopic mechanism to be responsible for these anomalies.
\end{abstract}


\maketitle

\draft

\vspace{2mm}

\end{titlepage}

Complex perovskite alloys have been studied for more than fifty
years, which are of great importance not only because they have
wildly technical applications for their large electromechanical
responses,\cite{KU} but also because they are fundamentally
interesting for their anomalous properties and the relations to
the microscopic structure. Recently, two breakthroughs were
achieved in the research of complex perovskites. One is the
observation of large enhancement of piezoelectric constants in
single crystals.\cite{Park,Serv} The second is the discovery of a
low-temperature monoclinic phase in Pb(Zr$_{1-x}$Ti$_{x}$)O$_{3}$
(PZT) near the morphotropic phase boundary (MPB).\cite{BN1,BN1a}
Inspired by such progresses, lots of works were done to
search larger electromechanical response and unobserved phases in
complex perovskites. Particularly, an \textit{ab initio} calculation
on Pb(Sc$_{0.5}$Nb$_{0.5}$)O$_{3}$ (PSN) revealed that atomic rearrangement
would lead simultaneously to large
electromechanical responses and unusual structural
phases.\cite{AM} The difference between the valences of the B
atoms (Sc$^{3+}$ and Nb$^{5+}$) was pointed out to be the main reason for the
existence of these anomalous properties and this mechanism is
expected to be generally applicable in a given class of
heterovalent perovskite alloys, which is likely to have large
technological and fundamental implications. However, it is unclear
whether there is any interesting effect in the modulated
homovalent systems where the B atoms have the same valences.

In this letter, we investigate the influences of compositional
modulation on the homovalent PZT solid solutions near MPB.
Remarkably, it is found that the atomic ordering along the [100] direction
 leads to the suppression of $P_{x}$ ($x$%
-component of the polarization). Unexpected new ground states of
triclinic (Tri) and C-type monoclinic (M$_{C}$)
phases appear with greatly enhanced
piezoelectric coefficients. The orientation dependence of
dipole-dipole interaction in modulated structure is revealed to
be responsible for these anomalies. Such microscopic mechanism is
expected to work in MPB region of any homovalent perovskite
when the atomic ordering is produced.

We adopt the effective Hamiltonian of PZT alloys proposed by
Bellaiche, Garcia and Vanderbilt, \cite{LB1} which is derived from
first-principles calculation, to predict the properties of
compositionally modulated PZT structures by conducting Monte Carlo
simulations. In this scheme, the total
energy $E$ is written as the sum of an average energy and a local energy as%
\cite{LB1,LB2}
\begin{eqnarray}
E(\{\mathbf{u}_{i}\},\{\mathbf{v}_{i}\},\eta _{H},\{\sigma _{j}\}) &=&E_{%
\mathrm{ave}}(\{\mathbf{u}_{i}\},\{\mathbf{v}_{i}\},\eta _{H})  \nonumber \\
&&+E_{\mathrm{loc}}(\{\mathbf{u}_{i}\},\{\mathbf{v}_{i}\},\{\sigma _{j}\}),
\end{eqnarray}%
where $\mathbf{u}_{i}$ is the local soft mode in unit cell $i$, $\mathbf{v}%
_{i}$ is the dimensionless local displacement related to the
inhomogeneous strain\cite{WZ}, $\eta _{H}$ is the homogeneous
strain tensors and $\sigma _{j}=\pm 1$ represents the presence of
a Zr or Ti atom, respectively, at lattice site $j$ of the PZT
alloy. All the parameters of Eq.~(1) are derived from the first
principle calculations and are listed in references.\cite{LB1,LB2}
The modulated structures under
consideration are made of the sequences Pb(Zr$_{1-x+\nu }$Ti$_{x-\nu }$)O$%
_{3}$/Pb(Zr$_{1-x-\nu }$Ti$_{x+\nu }$)O$_{3}$ along the [100]
direction with B atoms randomly distributing within each [100]
plane. 10$\times $10$\times $10 or 12$\times $12$\times $12
supercells with periodic boundary conditions are used in Monte
Carlo simulations to get well converged results. Eq.~(1) is based
on the virtual crystal approximation (VCA).\cite{No,LB4} The
parameters of Zr and Ti atoms are composition-dependent, which
reflects an average influence from surrounding atoms. Considering
that the modulation length of our studied structures is small
(only two monolayers) and a Zr or Ti atom will be influenced by
atoms from planes with different compositions, we adopt the
parameters corresponding to the overall composition $x$ for all
atoms in the modulated structures. Some detailed tests are also
carried out, which show that the anomalous properties of the
modulated structures revealed in this paper are not sensitive to
the choice of parameters.

\begin{figure}[tbp]
\includegraphics[width=8.5cm]{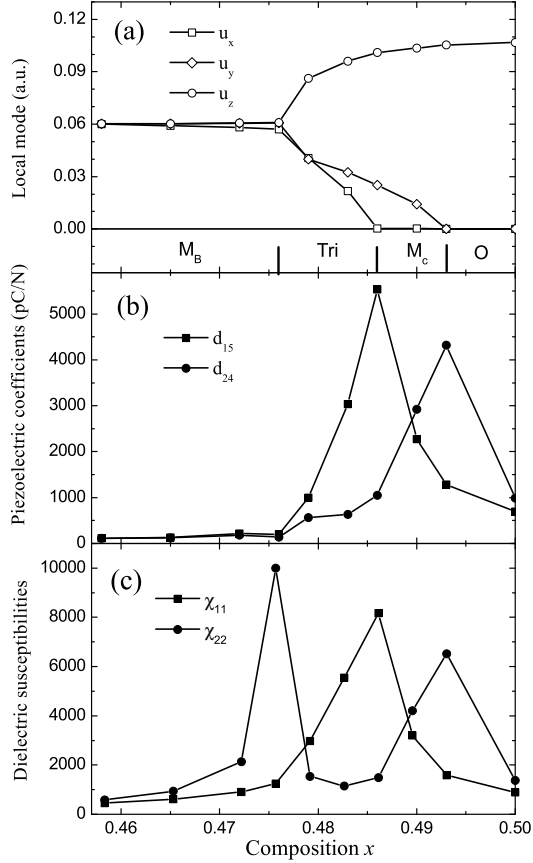}
\caption{ (a) Average cartesian coordinates of the local mode
vectors; (b) piezoelectric coefficients $d_{15}$ and $d_{24}$; (c)
dielectric susceptibilities $\chi_{11}$ and $\chi_{22}$ as
functions
of composition $x$ in atomic-ordered PZT alloys made of the Pb(Zr$_{1-x+\protect%
\nu}$Ti$_{x-\protect\nu}$)O$_{3}$/Pb(Zr$_{1-x-\protect\nu}$Ti$_{x+\protect\nu%
}$)O$_{3}$ sequences. $\nu$ is set to its maximal value as $\protect\nu=x$. The
temperature in simulation is 50K. }
\label{fig01}
\end{figure}

First we consider the properties of modulated PZT as a function of
composition $x$ by setting parameter $\nu$ to its maximum value
($\nu=x$). Fig.~1(a) shows the cartesian coordinates
($u_{x},u_{y},u_{z}$) of
the supercell average of the local mode vectors at $T$%
=50K. (Due to the difference between the theoretical and
experimental temperatures,\cite{LB1} it corresponds to an
\textquotedblleft experimental'' temperature around 30K.) In the
non-modulated case, it was demonstrated that disordered
Pb(Zr$_{1-x}$Ti$_{x}$)O$_{3}$ has a pseudocubic [111] direction
polarized ($u_{x}=u_{y}=u_{z}>0$) rhombohedral phase (R) for
$x<0.475$, a [001] direction polarized ($u_{x}=u_{y}=0<u_{z}$)
tetragonal phase (T) for $x>0.49$, and a A-type monoclinic phase
(M$_{A}$) with $0<u_{x}=u_{y}<u_{z}$ for
$0.475<x<0.49$.\cite{LB1,LB2} In the modulated case, for
compositions outside MPB, the local mode vectors are slightly
affected as shown in Fig.~1(a). However, the precise symmetry has
been changed by the atomic ordering, and the phases are recognized
as B-type monoclinic (M$_{B}$) and orthorhombic (O). The
interesting phenomena appear within the MPB region. For
$0.480<x<0.485$, $u_{x}$ decreases drastically, which results in a
triclinic (Tri) phase with $u_{x}\neq u_{y}\neq u_{z}$. When $x$
is larger than 48.5\%, $u_{x}$ vanishes and an unexpected ground
state appears in the modulated systems: a C-type monoclinic
(M$_{C}$) phase with $0=u_{x}<u_{y}<u_{z}$. Fig.~1(a) thus depicts
a polarization rotation in modulated PZT structures following the
sequence of M$_{B}$-Tri-M$_{C}$-O as the Ti composition increases.
Although Tri and M$_{C}$ phases in PZT were first demonstrated in
the simulation of electric-field induced transformation in
disordered rhombohedral PZT by Bellaiche \textit{et
al.},\cite{LB3} it has not been observed experimentally yet
because the predicted electric field required to induce the Tri
and M$_{C}$ phases is too high ($\sim$ 500 kV/cm for composition
close to MPB) compared with what experiments can provide. Thus the
present study demonstrates an alternative routine to observe such
phases with less experimental obstacles.

Fig.~1(b) shows the corresponding piezoelectric coefficients and
Fig.~1(c) the dielectric susceptibilities which are calculated in
the correlation-function approach\cite{AG}. It shows that $d_{15}$
reaches its maximum near the Tri to M$ _{C}$ transition and
remains a remarkably large value above 1000 pC/N in MPB region,
while $d_{24}$ achieves values above 1000pC/N in M$_{C}$ phase and
reaches its maximum near the M$_{C} $ to O transition. Another
important piezoelectric coefficient, $d_{33}$, is not shown here
because the modulations scarcely affect $u_{z}$ and have little
effect on $d_{33}$. The existences of the M$_{B}$-Tri, Tri-M$_{C}$
and M$_{C}$-O transitions also result in the peaks of dielectric
responses, as shown in Fig.~1(c).

\begin{figure}[tbp]
\includegraphics[width=8.5cm]{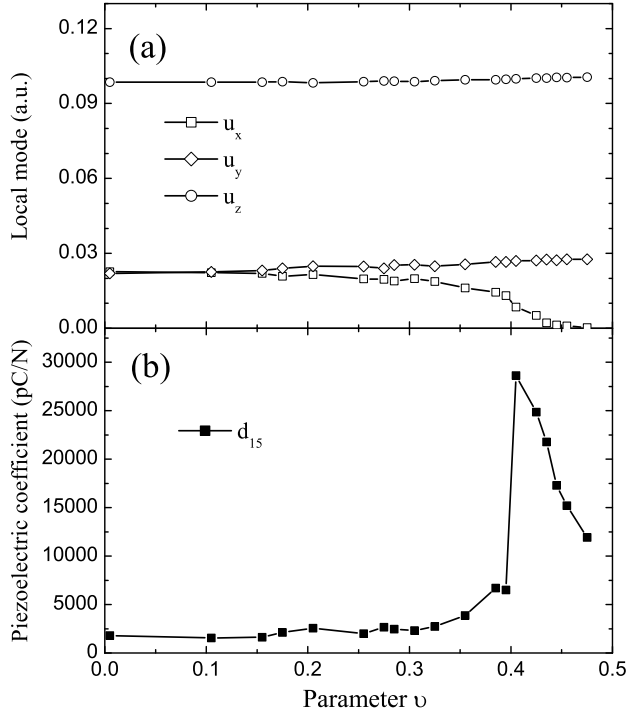}
\caption{ (a) Average cartesian coordinates of the local mode, and
(b) piezoelectric coefficient $d_{15}$, as functions of the parameter $%
\protect\nu$ in modulated PZT structures with $x=0.485$ at 50K. }
\label{fig02}
\end{figure}

We also investigate the effects of the modulation parameter $\nu$,
the fluctuation amplitude, on the structural and piezoelectric
properties of the modulated structures which have overall
composition $x=0.485$. Fig.~2(a) shows the average local modes as
a function of $\nu$ at 50K. The structure of $\nu=0$ (in the case
of disordered PZT alloy) has $u_{x}=u_{y}<u_{z}$ characterizing
the M$_{A}$ structure with polarization lying between [111] and
[001] directions. When $\nu$ becomes larger than 0.155, $u_{x}$
begins to decrease. With increasing $\nu$, $u_{x}$ decreases
gradually whereas $u_{y}$ and $u_{z}$ increase very slowly. This
behavior corresponds to the phase with a Tri symmetry. When $\nu$
becomes larger than 0.45, $u_{x}$ becomes null, while $u_{y}$ and
$u_{z}$ remain unequal, indicating a M$_{C}$ symmetry with a
polarization between [011] and [001] directions. Summarily,
increasing the parameter $\nu$ leads to a continuous rotation of
the polarization and transformation between different ground
states which are driven by the annihilation of $u_{x}$---the
polarization component along the direction of compositional
modulation.

Theoretical and experimental studies have suggested that polarization
rotation is essential for the large piezoelectric response in the perovskite
alloys.\cite{HF,BN3} Such an effect is clearly demonstrated in Fig.~2(b) where the
piezoelectric coefficient $d_{15}$ is depicted. With decreasing $u_{x}$,
$d_{15}$ increases from 2000 pC/N to 8000 pC/N gradually in the Tri phase for
$0.2<\nu<0.4$. Then $d_{15}$ reaches a very steep peak about 30000 pC/N at $%
\nu\simeq 0.4$ where the polarization most easily rotates [see
also Fig.~2(a)]. After that, $d_{15}$ decreases slowly but still
remains large values even in M$_{c}$ phase. As one can see in
Fig.~2(b), $d_{15}$ is larger than 10000 pC/N for a broad range of
$\nu>0.4$. To our best knowledge, such a huge enhancement of the
piezoelectric coefficient has never been reported in any
perovskite material before.

\begin{figure}[]
\includegraphics[width=8.5cm]{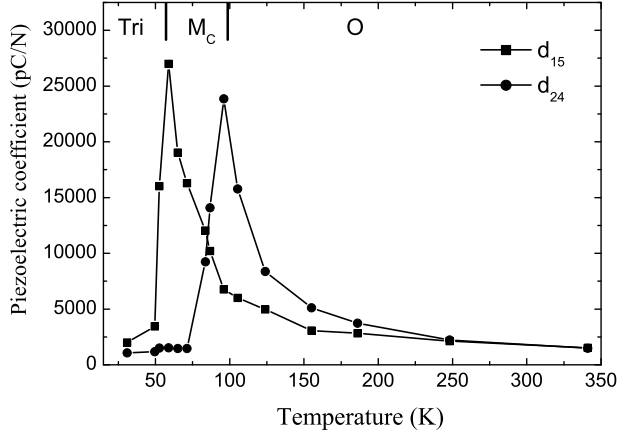}
\caption{ Piezoelectric coefficients in modulated PZT structure
with $x=0.485$ and $\protect\nu=0.305$ as functions of temperature. }
\label{fig03}
\end{figure}

Temperature properties of the material are important for its
application performances. Here we investigate the influence of
temperature on a structure with $x=0.485$ and $\nu=0.305$, which
adopts a Tri ground state in Fig.~2(a). The resulting
piezoelectric properties are depicted in Fig.~3. A huge $d_{24}$
piezoelectric coefficient peak appears near the O to M$_{C}$
transition (the information of average local mode is not shown
here), around which $d_{24}$ remains larger than 5000 pC/N for a
broad range of temperature. The simulations also predict the
$d_{15}$ peak consistent with the transition from M$_{C}$ to Tri
phase. In addition, it is found that with increasing $\nu$, the
M$_{C}$-Tri transition temperature decreases, and consequently,
the temperature of $d_{15}$ peak is dependent on the value of the
modulation parameter $\nu$. This dependence brings great
convenience to developing piezoelectric devices for application at
different temperatures.

The above intriguing results can be understood by
considering the orientation dependence of
dipole-dipole interaction in modulated structures.

Although PbZrO$_{3}$ and PbTiO$_{3}$ are similar enough to form Pb(Zr$_{1-x}$%
Ti$_{x}$)O$_{3}$ solid solutions at any composition $x$, their intrinsic
polarization properties are different to some extent.
As a result, the magnitude of
polarization vibrates with varying local composition that is caused by
the structural modulation. In our calculation, the Ti-rich planes
have larger average local mode, while Ti-poor planes have smaller values of
local mode.

\begin{figure}[]
\includegraphics[width=8.5cm]{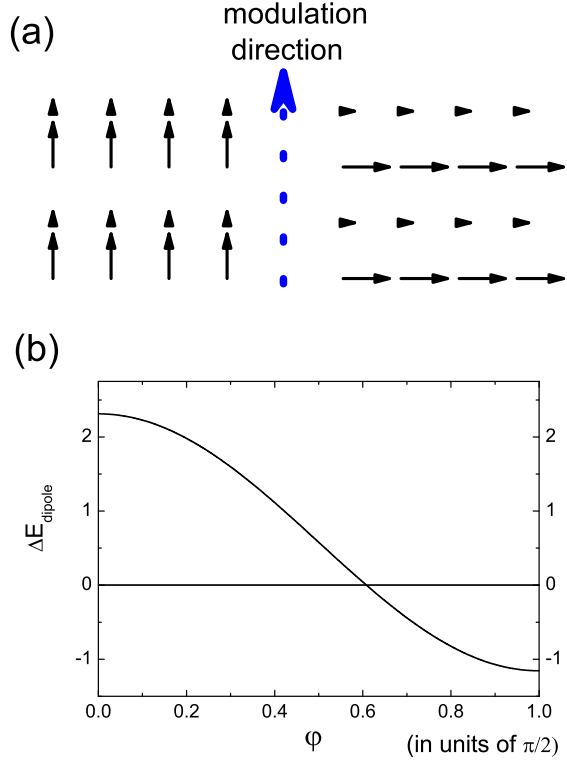}
\caption{ (a) Schematic graphics of dipole alignment in modulated structure:
dipole aligned parallel to modulation direction with higher energy (the left), and
alignment perpendicular to modulation with lower energy (the right). (b) Variation of
dipole-dipole interaction, $\Delta E_{\mathrm{dipole}} $ [see Eq.~(4)], as a
function of $\protect\varphi$ (the angle between the polarization and the
modulation direction). $\Delta E_{\mathrm{dipole}}$ is measured in units of $%
E_{\mathrm{dipole}}^0$, while $\protect\varphi$ in units of $\protect\pi/2$. }
\label{fig05}
\end{figure}

In the homovalent alloys such as PZT, the long-range polarization
interactions are described in dipole-dipole interaction energy:\cite{WZ,LB2}
\begin{equation}
E_{\mathrm{dipole}}=\frac{Z^{\ast 2}}{\epsilon _{\infty }}\sum_{i<j}\frac{%
\mathbf{u}_{i}\cdot \mathbf{u}_{j}-3(\mathbf{\hat{R}}_{ij}\cdot \mathbf{u}%
_{i})(\mathbf{\hat{R}}_{ij}\cdot \mathbf{u}_{j})}{R_{ij}^{3}},
\end{equation}%
where $Z^{\ast }$ is the Born effective charge and $\epsilon
_{\infty }$ the optical dielectric constant of the material. When
the dipole is uniform ($\mathbf{u}_{i}\equiv \langle
\mathbf{u}\rangle $), $E_{\mathrm{dipole}}$ is isotropic, \textit{
i.e.}, $E_{\mathrm{dipole}}$ is independent of the direction of
the polarization. However, when the magnitude of the dipole
fluctuates due to the structural modulation, the situation
differs. $E_{\mathrm{dipole}}$ becomes dependent on the angle
between the directions of modulation and polarization [see
Fig.~4(a)]: when the polarization is parallel to the modulation,
the dipoles are aligned shoulder to shoulder and the energy is
higher; when the polarization is perpendicular to the modulation,
the dipoles are aligned end to end so that the energy is lower. To
further clarify this point, we present a simple numerical analysis
here. The local modes in a [100] plane are assumed to fluctuate as
\begin{equation}
\mathbf{u}_{i}=\langle \mathbf{u}\rangle \lbrack 1+\xi K_{i}],
\end{equation}%
where $\xi $ is the fluctuation amplitude, and $K_{i}$ equals to $+1$ or $-1$
depending on which [100] plane the dipole locates. It is easy to show that
the dipole-dipole interaction energy can be rewritten as
\begin{equation}
E_{\mathrm{dipole}}=E_{\mathrm{dipole}}^{0}+\xi ^{2}\Delta E_{\mathrm{dipole}%
}.
\end{equation}%
$\Delta E_{\mathrm{dipole}}$ reflects the variation of dipole
energy caused by the structural modulation and relies on the angle
between the polarization and the modulation directions (denoted as
$\varphi $). In Fig.~4(b), we plot $\Delta E_{\mathrm{dipole}}$ as
a function of $\varphi $ with polarization rotating from [100] to
[010] direction. It shows that $\Delta E_{\mathrm{dipole}}$ has
its maximum when the polarization is parallel to the modulation
direction. When the
direction of polarization deviates from that of modulation, $\Delta E_{%
\mathrm{dipole}}$ continuously decreases and reaches the minimum
with polarization perpendicular to the modulation direction.
Accordingly, compositional modulation will cause the polarization
to rotate away from the modulation direction, which results in
unusual structure transitions and induces huge enhancement of
piezoelectric coefficient, as revealed in the above case of PZT.
In perovskite ultrathin films, the polarization is restricted
in the normal direction by the strain from matrix,
so the dipole-dipole interaction increases and violates the ferroelectricity
below a critical thickness.\cite{JJ}

It is noted that the structures outside MPB region are not
sensitive to the atomic ordering. The main reason is that the
dipole-dipole interaction is not the only part to determine the
Hamiltonian. When the rotation occurs, the other parts of the
Hamiltonian may increase and cause an energy barrier. MPB is a
region where different phases have close free energy, which means
that the energy barrier of polarization rotation is small. So the
dipole-dipole energy dominates in this region and produces the
anomalies. When the composition deviates from MPB, such energy
barrier increases and the polarization rotation is prevented.

On the basis of the general microscopic mechanism discussed here,
it is expected that solid solution made of homovalent perovskite
ferroelectrics with a MPB should have the similar structural and
piezoelectric anomalies as those of PZT when the atomic ordering
is arranged. Besides, compared with heterovalent alloys such as
PSN, experimental modifications to the atomic ordering of
homovalent alloys such as PZT is more practicable because
Pb(Zr$_{1-x}$Ti$_{x} $)O$_{3}$ is stoichimetric at any $x$ value.

This work was supported by State Key Program of Basic Research Development
(Grant No.~TG2000067108), the National Natural Science Foundation of China,
and Trans-century Training Programme Foundation for
the Talents by the Ministry of Education of China. W.~D.~acknowledges with
thanks the support of the Berkeley Scholar Program.

\end{document}